\documentstyle[preprint,aps,epsfig]{revtex}

\def\be{\begin{equation}}
\def\ee{\end{equation}}

\begin{document}
\draft

\title{Multipole polarizability of a graded spherical particle}

\date{\today}
\author{L. Dong$^{1,2}$, J. P. Huang$^{1,3}$,
 K. W. Yu$^{1,4}$\footnote{Corresponding author.
 Electronic mail: kwyu@phy.cuhk.edu.hk},
 G. Q. Gu$^{1,5}$}

\address{$^1$Department of Physics, The Chinese University of Hong Kong,
 \\ Shatin, New Territories, Hong Kong \\
 $^2$Biophysics and Statistical Mechanics Group,
 Laboratory of Computational Engineering, \\ Helsinki University of Technology,
 P.\,O. Box 9203, FIN-02015 HUT, Finland \\
 $^3$Department of Physics, Fudan University, Shanghai 200433, China \\
 $^4$Institute of Theoretical Physics, The Chinese University of Hong Kong,
 \\ Shatin, New Territories, Hong Kong \\
 $^5$College of Information Science and Technology,
 East China Normal University, \\ Shanghai 200 062, China
}
\maketitle

\begin{abstract}
We have studied the multipole polarizability of a graded spherical
particle in a nonuniform electric field, in which the conductivity can
vary radially inside the particle.
The main objective of this work is to access the effects of multipole
interactions at small interparticle separations, which can be important
in non-dilute suspensions of functionally graded materials.
The nonuniform electric field arises either from that applied on the
particle or from the local field of all other particles.
We developed a differential effective multipole moment approximation
(DEMMA) to compute the multipole moment of a graded spherical particle
in a nonuniform external field. Moreover, we compare the DEMMA results
with the exact results of the power-law graded profile and the agreement
is excellent. The extension to anisotropic DEMMA will be studied in an
Appendix.
\end{abstract}
\vskip 5mm

\section{Introduction}

In functionally graded materials (FGM), the materials or microstructure
properties may vary spatially to meet the specific needs in various
engineering applications \cite{Yamanouchi}.
The material or microstructure gradients in FGM lead to properties quite
distinct from those of the homogeneous materials and conventional
composite materials \cite{Yamanouchi,Holt}.
A unique advantage of using FGM is that one can tailor the materials
properties via the design of the gradients. Over the past few years,
there have been overwhelming attempts, both theoretical and experimental,
of studying the responses of FGM to mechanical, thermal, and electric
loads and for different microstructure in various systems
\cite{Yamanouchi,Holt,Ilschner,fgm1,fgm2,fgm3,fgm4,fgm5}.
Such FGM may exist in Nature or they can be made artificially.
For instance, graded morphogen profiles can exist in a cell layer
\cite{fgm1}.
Graded structure may be produced by using various approaches, such as a
three-dimensional X-ray microscopy technique \cite{fgm4}, deformation
under large sliding loads \cite{fgm5}, and adsorbate-substrate atomic
exchange during growth \cite{fgm3}.
Moreover, it has been reported recently that the control of a
compatibility factor can facilitate the engineering of FGM \cite{fgm2}.

There have been various theoretical attempts to treat the composite
materials of homogeneous inclusions \cite{Jackson} as well as multi-shell
inclusions \cite{Gu1,Fuhr,Arnold,Chan}. These established theories for
homogeneous inclusions, however, cannot be applied to graded inclusions
directly.
To this end, we have recently developed a first-principles approach for
calculating the effective response of dilute composites of graded
cylindrical inclusions \cite{GuYu} as well as graded spherical particles
\cite{Dong2003}. More recently, the spectral representation of graded
composites has also been established \cite{Dong2005}.
These theories essentially assumed that the particles are sufficiently
far apart so that it is possible to neglect contributions from
higher-order multipoles and these approaches are only valid for dilute
composites. As the particles become closer, the local field is
extremely inhomogeneous on the surface of the particles.
To this end, the electromagnetic response of a system of homogeneous
(non-graded) spherical particles has been studied extensively in the
literature \cite{Hulst,Rojas88}.
In this work, we aim at studying the multipole polarizability of a graded
spherical particle in a nonuniform field which occurs naturally for a
system of interacting particles. The electrostatic boundary-value problem
of a graded spherical particle will be solved to obtain exact analytic
results for the power-law graded profile. For arbitrary graded profiles,
we will develop a differential effective multipole moment approximation
(DEMMA) to compute the multipole moment of a graded spherical particle to
capture the multipole response of the particle in a nonuniform field.

The plan of the paper is as follows. In section II, we will solve the
boundary-value problem of a graded spherical particle in a nonuniform
electric field. In this way, the exact analytic expression for the
multipole
polarizability is obtained for the case of a power-law conductivity profile.
In section III, we derive the DEMMA for the multipole polarizability of a
graded spherical particle of an arbitrary graded profile. In section IV,
we compare the DEMMA results with the exact results of first-principles
approach. Discussion of our approaches will be given in section V.
The generalization to anisotropic DEMMA will be discussed.

\section{First-principles approach}

We consider a graded spherical inclusion of radius $a$ subjected to a
nonuniform electric field of a point charge placed on the $z$-axis.
For electrical conductivity, the constitutive relations read
$\vec{J}=\sigma_{i}(r)\vec{E}$ and $\vec{J}=\sigma_{m}\vec{E}$
respectively in the graded spherical inclusion and the host medium,
where $\sigma_{i}(r)$ is the conductivity profile of the graded spherical
inclusion and $\sigma_{m}$ is the conductivity of the host medium.
Moreover,
$$
\vec{\nabla} \cdot \vec{J}=0,\ \ \ \
\vec{\nabla} \times \vec{E}=0.
$$
To this end, $\vec{E}$ can be written as the gradient of a scalar
potential $\Phi$, $\vec{E}=-\vec{\nabla}\Phi$,
leading to a partial differential equation:
\be
\vec{\nabla} \cdot [\sigma(r)\vec{\nabla} \Phi] = 0.
\ee
where $\sigma(r)$ is the dimensionless conductivity profile, while
$\sigma(r)=\sigma_i(r)/\sigma_m$ in the inclusion, and $\sigma(r)=1$ in
the host medium. Without loss of generality, we may also let $a=1$.

In spherical coordinates, the electric potential $\Phi$ satisfies
\be
\frac{1}{r^2}\frac{\partial}{\partial r} \left(r^2\sigma(r)
\frac{\partial \Phi}{\partial r}\right)
+\frac{1}{r^2\sin\theta}\frac{\partial}{\partial \theta}
\left(\sin\theta \sigma(r) \frac{\partial \Phi}{\partial \theta}\right)
+\frac{1}{r^2\sin^2\theta}\frac{\partial}{\partial \varphi}
\left(\sigma(r)\frac{\partial \Phi}{\partial \varphi}\right)=0.
\label{comm}
\ee

We place a point charge of magnitude $q$ on the $z$-axis at a distance
$R$ from the center of the graded sphere.
Thus $\Phi$ is independent of the azimuthal angle $\varphi$. If we write
$\Phi = f(r)\Theta(\theta)$ to achieve separation of variables, we obtain
two distinct ordinary differential equations.
For the radial function $f(r)$,
\be
\frac{d}{dr}\left(r^2 \sigma(r) \frac{df}{dr}\right)-l(l+1)\sigma(r)f=0,
\label{general}
\ee
where $l$ is an integer, while $\Theta(\theta)$ satisfies the Legendre
equation \cite{Jackson}. The potential can be obtained by solving
Eq.(\ref{general}).
Exact analytic results can be obtained for a power-law profile
\cite{GuYu,Dong2003}, linear profile \cite{GuYu}, and exponential profile
\cite{Martin,Gu,Wei}.

Let us consider the case in which the conductivity profile of the particle
has a power-law dependence on the radius, $\sigma(r)=c r^k$, with
$k \ge 0$ where $0< r \le 1$. Then the radial equation becomes
\be
\frac{d^2 f}{dr^{2}}+\frac{k+2}{r}\frac{df}{dr}-\frac{l(l+1)f}{r^2}=0.
\label{Re}
\ee
As Eq.(\ref{Re}) is a homogeneous equation, it admits a power-law solution
\cite{Dong2003},
\be
f(r)=r^{s}.
\ee
\label{f(r)}
Substituting it into Eq.(\ref{Re}), we obtain the equation
$s^{2}+s(k+1)-l(l+1)=0$
and the solution is
\be
s^{k}=\frac{1}{2}\left[-(k+1)\pm\sqrt{(k+1)^{2}+4l(l+1)} \right].
\label{sp}
\ee

The potentials in the inclusion and the host medium are given,
respectively, by
\begin{eqnarray}
\Phi_{i}(r,\theta)&=& \sum_{l=0}^{\infty} A_{l}
r^{s_{+}^{k}(l)} P_{l}(\cos\theta),\nonumber\\
\Phi_{m}(r,\theta)&=&\frac{q}{r_q}+\sum_{l=0}^{\infty} B_{l}
r^{-(l+1)} P_{l}(\cos\theta).
\label{potential}
\end{eqnarray}
The positive root of Eq.(\ref{sp}) has been chosen for nonsingular
solution inside the inclusion.
Meanwhile, the potential functions should satisfy the boundary
conditions, as follow:
\begin{eqnarray}
\Phi_{i}(r,\theta)\left|_{r=1} \right.&=&\Phi_{m}(r,\theta)\left|_{r=1}
\right.,\nonumber\\
\sigma(r)\frac{\partial \Phi_{i}(r,\theta)}{\partial r}\left|_{r=1}
\right.&=&\frac{\partial \Phi_{m}(r,\theta)}{\partial r}\left|_{r=1}
\right..
\label{bc}
\end{eqnarray}
The potential due to a point charge located at a distance $R$ from the
center of the particle can be rewritten in a multipole expansion in the
vicinity of the particle surface \cite{Jackson}:
\be
\frac{q}{r_q}=q\sum_{l=0}^{\infty} \frac{r^{l}}{R^{l+1}}P_{l}(\cos\theta).
\ee
From Eq.(\ref{bc}), we obtain a set of simultaneous linear equations
$$
A_{l}=\frac{q}{R^{l+1}}+B_{l},
\ \ \ \
A_{l}=\frac{l \sigma_{m}}{\sigma_{i}(1) s_{+}^{k}(l)}
\left(\frac{q}{R^{l+1}}-\frac{l+1}{l} B_{l} \right).
$$

Solving the above equations, we obtain the coefficients:
\begin{eqnarray}
A_{l}&=&\frac{q}{R^{l+1}}\frac{2l+1}{l(F_{l}+1)+1},\nonumber\\
B_{l}&=&-\frac{q}{R^{l+1}}\frac{l(F_{l}-1)}{l(F_{l}+1)+1},
\ \ \ l\ge 1,
\end{eqnarray}
where
\be
F_l=\frac{\sigma_{i}(1)}{\sigma_m}\frac{s_{+}^{k}(l)}{l}, \ \ \ l\ge 1.
\label{Fl}
\ee

The coefficients $B_l$ are just proportional to the multipole response
of the graded spherical particle to the applied multipole field.
We thus identify the multipole factor
\be
H_l=\frac{l(F_{l}-1)}{l(F_{l}+1)+1}, \ \ \ l\ge 1.
\label{exact}
\ee

From Eq.(\ref{Fl}) and Eq.(\ref{exact}), the quantity $F_l$ can be
identified as the $l$-dependent equivalent conductivity of the graded
particle [see Eq.(\ref{eqiv}) in section III below].
In the case of a homogeneous sphere, $k=0$, $s_+(l)=l$,
$F_l=\sigma_i/\sigma_m$, one recovers the well known result \cite{Hulst}:
$$
H_l=\frac{l(\sigma_i-\sigma_m)}{l(\sigma_i+\sigma_m)+\sigma_m,},
\ \ \ l\ge 1.
$$
For a uniform field, however, the dipole factor of Dong et al
\cite{Dong2003} recovers.

\section{Differential effective multipole moment approximation}

In this section, we develop the differential effective multipole moment
approximation (DEMMA) for a graded spherical particle and hence compute
the multipole factor. To establish the differential effective multipole
moment theory, we mimic the graded profile by a multi-shell construction
\cite{Yu02}, i.e., we build up the conductivity profile gradually by
adding shells. We start with an infinitesimal spherical core of
conductivity $\sigma_i(0^+)$ and keep on adding spherical shells of
conductivity given by $\sigma_i(r)$ at radius $r$, until $r=a$ is
reached. At radius $r$, we have an inhomogeneous sphere with certain
multipole moment. We further replace the inhomogeneous sphere by a
homogeneous sphere of the same multipole moment and the graded profile is
replaced by an effective conductivity $\bar{\sigma}_i(r)$. Thus,
\be
H_l(r)=\frac{l(\bar{\sigma}_i(r)-\sigma_m)}
{l(\bar{\sigma}_i(r)+\sigma_m) + \sigma_m}.
\label{Hlr}
\ee
Next, we add to the sphere a spherical shell of infinitesimal thickness
${\rm d}r$, of conductivity $\sigma_i(r)$. The resulting multipole factor
$H_{l}'$ will change according to~\cite{Rojas88}
\be
H_{l}' = \frac{(1-\sigma_m/\sigma_i(r))\sigma_i(r)
(l'\sigma_i(r)+l\bar{\sigma}_i(r))+\rho \sigma_i(r)
[l'(\bar{\sigma}_i(r)-\sigma_i(r))+l\sigma_m
(\bar{\sigma}_i(r)/\sigma_i(r)-1)]}
{[1+l'\sigma_m/(l\sigma_i(r))]\sigma_i(r)
(l'\sigma_i(r)+l\bar{\sigma}_i(r))
+l'\rho\sigma_i(r)(\sigma_m-\sigma_i(r)
+\bar{\sigma}_i(r)-\sigma_m\bar{\sigma}_i(r)/\sigma_i(r))}
\label{coated}
\ee
with $l'=l+1$ and $\rho = [r/(r+{\rm d}r)]^{2l+1}$.

Of course, the equivalent conductivity $\bar{\sigma}_i(r)$, being related
to $H_l(r)$, should also change by the same token.
Let us write $H'_l = H_l + {\rm d}H_l$, and take the limit ${\rm d}r\to 0$,
we obtain a differential equation:

\begin{eqnarray}
\frac{{\rm d}H_l(r)}{{\rm d} r}
&=&
-\frac{1}{(2l+1)r\sigma_m\sigma(r)}
[(H_l(r)+l+H_l(r)l)\sigma_m+(H_l(r)-1)l\sigma(r)] \nonumber \\
&\times& [(H_l(r)+l+H_l(r)l)\sigma_m-(H_l(r)-1)(l+1)\sigma(r)].
\label{DEMMA}
\end{eqnarray}
Thus the multipole factor of a graded spherical particle can be
calculated by solving the above differential equation with a given graded
profile $\sigma_i(r)$. Solving $H_l(r)$ from Eq.(\ref{DEMMA}) gives
numerical results for the multipole factor. The nonlinear first-order
differential equation can be integrated if we are given the graded
profile $\sigma_i(r)$ and the initial condition $H_l(r=0)$.
In general, it works for arbitrary graded profiles $\sigma_i(r)$.

The substitution of the relation Eq.(\ref{Hlr}) into Eq.~(\ref{DEMMA})
yields the differential equation for the equivalent conductivity
$\bar{\sigma}_i(r)$,
\be
\frac{{\rm d}\bar{\sigma}_i(r)}{{\rm d} r}  =
\frac{[\sigma_i(r)-\bar{\sigma}_i(r)]
[(l+1)\sigma_i(r)+l\bar{\sigma}_i(r)]}{r\sigma_i(r)}.
\label{eqiv}
\ee
The $l=1$ limit of Eq.~(\ref{eqiv}) is a special case of the Tartar
formula, derived for assemblages of spheres with varying radial and
tangential conductivities \cite{Milton}.

\section{Numerical Results}


Figure~\ref{comparison} displays the comparison between the DEMMA
[Eq.~(\ref{DEMMA})] and the first-principles approach
[Eq.~(\ref{exact})], for the model power-law graded profiles
$\sigma_i(r) = c r^k.$ Here $H_l$ is plotted as a function of $c$,
for different $k$. As $c$ increases, the multipole factor $H_l$ is caused
to increase monotonically, for different multipole order $l$. The effect
of $l$ can also increase the $H_l$ slightly.
Nevertheless, increasing $k$ causes the $H_l$ to decrease.
Interestingly, the excellent agreement between the two methods have been
shown in this figure. In fact, the DEMMA is valid for arbitrary gradation
profiles, as mentioned above. However, exact analytic results obtained from
the first-principles approach often lack except for a few specific
graded profiles like the power-law profiles. In view of the good
comparison shown in this figure, one can safely use the DEMMA to treat
other graded profiles which might be not possible, or difficult, to
be solved by using the first-principles approach. To this end, the good
agreement between DEMMA results and the exact results can be understood
by the fact that Eq.(\ref{Fl}) indeed solves Eq.(\ref{eqiv}) for the
power-law profile. Moreover, the approximate DEMMA approach has been shown
to be exact for spherical particles \cite{YuGu2005}.

In Fig.~\ref{equiv_sigma}, we show the equivalent conductivity
$\bar{\sigma}_i(r=a)$ of a graded spherical particle, according to
the DEMMA [Eq.~(\ref{eqiv})], for the same model graded profile
$\sigma_i(r) = c r^k.$ Eq.~(\ref{eqiv}) shows that $\bar{\sigma}_i(r=a)$
depends only on the multipole order $l$ and the conductivity gradation
profile $\sigma_i(r)$. Here $a$ denotes the radius of the particle, which
has been set to be unity throughout the paper.  Similarly, increasing $c$
or $l$ causes the equivalent conductivity to increase. However, opposite
behavior is obtained for increasing $k$.

In fact, the parameter $k$ in the graded profile $\sigma_i(r)=c r^k$
measures the degree of inhomogeneity in the graded particle.
From Figs.~\ref{comparison}~and~\ref{equiv_sigma}, it is concluded that
the presence of inhomogeneity in the particle can affect the electrical
properties of the particles significantly. In other words, once an
inhomogeneous particle is used in reality, its effect of inhomogeneity
should be taken into account.

\section{Discussion and conclusion}

Here a few comments are in order. In this work, we have developed a
differential effective multipole moment approximation (DEMMA) to compute
the multipole moment of a graded spherical particle. We compared the DEMMA
results with the exact results of the power-law profile and the agreement
is excellent. Note that an exact solution is very few in composite
research and to have one yields much insight. Such solutions should be
useful as benchmarks.

We are now in a position to propose some applications of the present
theory. As the multipole response is sensitive to the graded profile
of the particles as well as to the structure of the nonuniform field
source, there is a potential application to ac electrokinetics of
graded colloidal particles \cite{Huang2004}. The similar approach can be
applied to electrorheological (ER) fluids \cite{jcp1,jcp2} because the
particles in ER fluids can have very dense structures locally.
The local electric fields are extremely inhomogeneous near the particles
so that multipole effects can play an important role. In this regard,
we can also study the interparticle force between graded particles
\cite{Yu} in ER fluids.

In the other topics, we may attempt the similar calculation of the
multipole response of a graded metallic sphere in the nonuniform field of
an oscillating point dipole at optical frequency. The graded Drude
dielectric function can be adopted \cite{APL}. When the oscillating
source is placed close enough to the graded metallic sphere, higher-order
multipole response can be excited.
The approach may also be applied to the electroencephalogram of the human
brain by regarding the brain as a graded anisotropic conducting sphere
\cite{Dong2004}. In this way, the forward problem can be analyzed.
The derivation of the anisotropic DEMMA will be given in Appendix A.

In summary, we have studied the multipole response of a graded spherical
particle in a nonuniform field. We developed a DEMMA to compute the
multipole moment of a graded spherical particle. Moreover, we compare the
DEMMA results with the exact results of the power-law profile and the
agreement is excellent.

\section*{Acknowledgments}

This work was supported by the RGC Earmarked Grant of the Hong Kong SAR
Government.
G. Q. G. acknowledges the support from the National Science Foundation
of China under grant number 10374026.
K. W. Y. acknowledges useful discussion with Joseph Kwok and C. T. Yam.

\newpage

\begin{appendix}

\section{Anisotropic differential effective multipole moment approximation}

Let us consider a graded spherical particle with radius $a$.
We adopt the spherical coordinates for convenience. The graded spherical
particle has a tangential conductivity in the plane orthogonal to the
radial vector of the sphere $\sigma_i^{\bot}(r)$ and a radial
conductivity $\sigma_i{}^{\|}(r)$. Both $\sigma_i^{\bot}(r)$ and
$\sigma_i{}^{\|}(r)$ will be prescribed by radial functions, $0< r \le
a$. In view of the symmetry, the anisotropic conductivity of the graded
sphere can be expressed as a tensor form
${\mathord{\buildrel{\lower3pt\hbox{$\scriptscriptstyle\leftrightarrow$}}\over
\sigma_i(r)}}$, namely,
\begin{equation}
{\mathord{\buildrel{\lower3pt\hbox{$\scriptscriptstyle\leftrightarrow$}}\over
\sigma_i(r)}} =
\sigma_i{}^{\|}(r) \hat{r}\hat{r} +
\sigma_i{}^{\bot}(r) \hat{\theta}\hat{\theta} +
\sigma_i{}^{\bot}(r) \hat{\varphi}\hat{\varphi}.
\end{equation}

Next, we present an anisotropic differential effective multipole moment
approximation (ADEMMA), which, similar to DEMMA, is a numerical method for
the analysis of the electric property of anisotropic graded particles
with arbitrary gradation profiles. Similarly, we may regard the gradation
profile as a multishell construction. In detail, we establish the
electric profile gradually by adding shells. Let us start with an
infinitesimal isotropic spherical core with conductivity $\sigma_i(0^+)$,
and keep on adding shells with both tangential and normal conductivity
profiles $\sigma_i^{\bot}(r)$ and $\sigma_i{}^{\|}(r)$ at radius $r$,
until $r=a$ is reached. At radius $r$, we have an inhomogeneous particle,
and further regard such an inhomogeneous particle as an effective
homogeneous one with an effective conductivity $\bar{\sigma}_i(r)$, which
has the multipole factor
\begin{equation}
H_l(r) = \frac{l (\bar{\sigma}_i(r)-\sigma_m)}
{l(\bar{\sigma}_i(r)+\sigma_m)+\sigma_m}.
\label{effective}
\end{equation}
Then, we add to the particle a shell with infinitesimal thickness
${\rm d} r$, with conductivities  $\sigma_i{}^{\bot}(r)$ and
$\sigma_i{}^{\|}(r)$. Its multipole factor $H_l'$ should change
according to the multipole factor of a single-coated particle
\cite{Lucas}
\begin{equation}
H_l' = \frac{(\sigma_i{}^{\|}(r) u_- -\bar{\sigma}_i(r) l)
(\sigma_i{}^{\|}(r)u_+-\sigma_m l)-\rho_l (\sigma_i{}^{\|}(r)
u_+-\bar{\sigma}_i(r) l) (\sigma_i{}^{\|}(r) u_--\sigma_m l )
}{(\bar{\sigma}_i(r)l-\sigma_i{}^{\|}(r) u_+ ) (\sigma_i{}^{\|}(r)
u_-+ \sigma_m l')\rho_l-(\bar{\sigma}_i(r) l-\sigma_i{}^{\|}(r)
u_-) (\sigma_i{}^{\|}(r) u_++\sigma_m l') },
\end{equation}
with $u_{\pm} = [-1\pm \sqrt{ 1+4l (1+l)
\sigma_i^{\bot}(r)/\sigma_i{}^{\|}(r)} ]/2$, $\rho_l
= [r/(r+{\rm d}r)]^{u_+-u_-},$ and $l' = 1+l.$ Let us write further $H_l'
= H_l + {\rm d}H_l$, and take the limit ${\rm d}r\to 0,$ we obtain a
differential equation
\begin{eqnarray}
\frac{{\rm d}H_l(r)}{{\rm d}r} &=& -\frac{1}{ (1+2l)r\sigma_m \sigma_i{}^{\|}(r) } [l\sigma_m-u_- \sigma_i{}^{\|}(r) +H_l(r)(l'\sigma_m+u_-\sigma_i{}^{\|}(r) )] \nonumber\\
& &[l\sigma_m-u_+ \sigma_i{}^{\|}(r) +H_l(r) (l'\sigma_m+u_+ \sigma_i{}^{\|}(r))].
\label{ADEMMA}
\end{eqnarray}
Thus the multipole factor of an anisotropic graded spherical particle
$H_l(r=a)$ can be calculated by solving the nonlinear first-order
differential equation [Eq.~(\ref{ADEMMA})] which can be integrated, at
least numerically if we are given the graded profiles
$\sigma_i^{\bot}(r)$ and $\sigma_i{}^{\|}(r)$ and the initial condition
$H_l(r=0).$ The substitution of Eq.~(\ref{effective}) into
Eq.~(\ref{ADEMMA}) yields the differential equation for the equivalent
conductivity
\begin{equation}
\frac{{\rm d} \bar{\sigma}_i(r)}{{\rm d}r } = \frac{(1+l)
\sigma_i{}^{\|}(r)  \sigma_i{}^{\bot}(r)- \sigma_i{}^{\|}(r)
\bar{\sigma}_i(r) -l \bar{\sigma}_i(r)^2  }{r\sigma_i{}^{\|}(r)
}.
\label{generalized-Tartar}
\end{equation}
Eqs.~(\ref{ADEMMA})~and~(\ref{generalized-Tartar}) can respectively reduce
to Eqs.~(19)~and~(20), as long as there is
$\sigma_i^{\bot}(r) =\sigma_i{}^{\|}(r) = \sigma_i(r).$
The substitution of $l=1$ into Eq.~(\ref{generalized-Tartar}) reduces
to the Tartar formula, derived for assemblages of spheres with varying
radial and tangential conductivities \cite{Milton}.
Last but not least, it is also instructive to extend the first-principles
approach of Ref.\cite{Gu} to multipole response and compare with the
ADEMMA results.

\end{appendix}

\begin{figure}[h]
\caption{Comparison of multiple factor $H_l$, between the DEMMA
[Eq.~(\ref{DEMMA})] and the first-principles approach
[Eq.~(\ref{exact})]. The graded profile is $\sigma_i(r) = cr^k$ and
$\sigma_m=1.$
Excellent agreement between the two methods have been shown.}
\label{comparison}
\end{figure}

\begin{figure}[h]
\caption{The equivalent conductivity $\bar{\sigma}_i(r=a)$ of a graded
spherical particle, according to the DEMMA [Eq.~(\ref{eqiv})].
Here $a$ denotes the radius of the particle,
which is set to unity throughout the paper.
The graded profile is $\sigma_i(r) = c r^k.$ }
\label{equiv_sigma}
\end{figure}

\newpage
\centerline{\epsfig{file=f1.eps,width=250pt}}
\centerline{Fig.~1}

\newpage
\centerline{\epsfig{file=f2.eps,width=250pt}}
\centerline{Fig.~2}

\begin{references}

\bibitem{Yamanouchi} M. Yamanouchi, M. Koizumi, T. Hirai, and I. Shioda,
 in {\em Proceedings of the First International Symposium on
 Functionally Graded Materials},
 edited by M. Yamanouchi, M. Koizumi, T. Hirai, and I. Shioda
 (Sendi, Japan, 1990).
\bibitem{Holt} J. B. Holt, M. Koizumi, T. Hirai, and Z. A. Munir,
 Ceramic transaction: functionally graded materials, Vol. {\bf 34},
 Westerville, OH: The American Ceramic Society, 1993.
\bibitem{Ilschner} B. Ilschner and N. Cherradi,
 {\em Proceedings of the Third International Symposium on Structural and
 Functionally Graded Materials}, Lausanne, Switzerland:
 Presses Polytechniquies et Universititaires Romands, 1994.

\bibitem{fgm1} T. Bollenbach, K. Kruse, P. Pantazis, M. Gonzalez-Gaitan,
 and F. Julicher, Phys. Rev. Lett. {\bf 94}, 018103 (2005).
\bibitem{fgm2} G. J. Snyder and T. S. Ursell, Phys. Rev. Lett. {\bf 91},
 148301 (2003).
\bibitem{fgm3} D. S. Lin, J. L. Wu, S. Y. Pan, and T. C. Chiang,
 Phys. Rev. Lett. {\bf 90}, 046102 (2003).
\bibitem{fgm4} B. C. Larson, W. Yang, G. E. Ice, J. D. Budai,
 and J. Z. Tischler, Nature (London) {\bf 415}, 887 (2002).
\bibitem{fgm5} D. A. Hughes and N. Hansen, Phys. Rev. Lett. {\bf 87},
 135503 (2001).

\bibitem{Jackson} J. D. Jackson, {\em Classical Electrodynamics}
 (Wiley, New York, 1975).
\bibitem{Gu1} G. Q. Gu and K. W. Yu, Acta Physica Sinica {\bf 40},
 709 (1991).
\bibitem{Fuhr} G. Fuhr and P. I. Luzmin, Biophys. J. {\bf 50}, 789 (1986).
\bibitem{Arnold} W. M. Arnold and U. Zimmermann, Z. Naturforsch.
 {\bf 37c}, 908 (1982).
\bibitem{Chan} K. L. Chan, P. R. C. Gascoyne, F. F. Becker, and P. Pethig,
 Biochim. Biophys. Acta {\bf 1349}, 182 (1997).
\bibitem{GuYu} G. Q. Gu and  K. W. Yu, J. Appl. Phys. {\bf 94}, 3376 (2003).
\bibitem{Dong2003} L. Dong, G. Q. Gu, and K. W. Yu, Phys. Rev. B
 {\bf 67}, 224205 (2003).
\bibitem{Dong2005} L. Dong, M. Karttunen, and K. W. Yu, Phys. Rev. E
 {\bf 72}, 016613 (2005).

\bibitem{Hulst} H. C. Van de Hulst, {\em Light Scattering by Small Particles}
 (Dover, New York, 1981).
\bibitem{Rojas88} R. Rojas, F. Claro, and R. Fuchs, Phys. Rev. B {\bf 37},
 6799 (1988).

\bibitem{Martin} P. A. Martin, J. Eng. Math. {\bf 42}, 133 (2002).
\bibitem{Gu} G. Q. Gu and  K. W. Yu, J. Compos. Mater. {\bf 39}, 127 (2005).
\bibitem{Wei} E. B. Wei, J. B. Song, and J. W. Tian, Phys. Lett. A
 {\bf 319}, 401 (2004).
\bibitem{Yu02} K. W. Yu, G. Q. Gu, and J. P. Huang, cond-mat/0211532.

\bibitem{Milton} G. W. Milton, {\em The Theory of Composites},
 (Cambridge University Press, Cambridge, 2002), p.~121.

\bibitem{YuGu2005} K. W. Yu and G. Q. Gu, Phys. Lett. A {\bf 345},
 448 (2005); cond-mat/0507325.
\bibitem{Huang2004} J. P. Huang, Mikko Karttunen, K. W. Yu, L. Dong, and
 G. Q. Gu, Phys. Rev. E {\bf 69} 051402 (2004).
\bibitem{jcp1} J. P. Huang and K. W. Yu,
 J. Chem. Phys. {\bf 121}, 7526 (2004).
\bibitem{jcp2} G. Q. Gu, K. W. Yu and P. M. Hui,
 J. Chem. Phys. {\bf 116}, 10989 (2002).
\bibitem{Yu} K. W. Yu and J. T. K. Wan, Comput. Phys. Commun. {\bf 129},
 177 (2000).

\bibitem{APL} J. P. Huang and K. W. Yu, Appl. Phys. Lett. {\bf 85},
 94 (2004).
\bibitem{Dong2004} L. Dong, J. P. Huang, K. W. Yu, and G. Q. Gu,
 J. Appl. Phys. {\bf 95}, 621 (2004).

\bibitem{Lucas} A. A.  Lucas, L. Henrard, and P. Lambin, Phys. Rev. B
 {\bf 49}, 2888 (1994).

\end{references}
\end{document}